\documentclass{jfm}

\usepackage{amsmath,amssymb}
\usepackage{graphicx}
\usepackage{natbib}
\usepackage{enumitem}
\usepackage{cleveref}
\usepackage{csquotes}
\usepackage{mathmacros}

\ifCUPmtlplainloaded \else
  \checkfont{eurm10}
  \iffontfound
    \IfFileExists{upmath.sty}
      {\typeout{^^JFound AMS Euler Roman fonts on the system,
                   using the 'upmath' package.^^J}%
       \usepackage{upmath}}
      {\typeout{^^JFound AMS Euler Roman fonts on the system, but you
                   dont seem to have the}%
       \typeout{'upmath' package installed. JFM.cls can take advantage
                 of these fonts,^^Jif you use 'upmath' package.^^J}%
      }
  \else
  \fi
\fi

\ifCUPmtlplainloaded \else
  \checkfont{msam10}
  \iffontfound
    \IfFileExists{amssymb.sty}
      {\typeout{^^JFound AMS Symbol fonts on the system, using the
                'amssymb' package.^^J}%
       \usepackage{amssymb}%
         \let\leq=\leqslant
       \let\ge=\geqslant  
      }{}
  \fi
\fi

\ifCUPmtlplainloaded \else
  \IfFileExists{amsbsy.sty}
    {\typeout{^^JFound the 'amsbsy' package on the system, using it.^^J}%
     \usepackage{amsbsy}}
    {\providecommand\boldsymbol[1]{\mbox{\boldmath $##1$}}}
\fi

\providecommand\bnabla{\boldsymbol{\nabla}}

\newsavebox{\astrutbox}
\sbox{\astrutbox}{\rule[-5pt]{0pt}{20pt}}

\usepackage{mathtools}
\DeclarePairedDelimiter\abs{\lvert}{\rvert}
\renewcommand{\vec}[1]{\ensuremath{\mathbf{#1}}}
\newcommand{\fourier}[1]{\ensuremath{\hat{#1}}}
\newcommand{\pdual}[1]{\ensuremath{\hat{#1}}}
\newcommand{\wavevecset}[1]{\mathcal{#1}}
\newcommand{\domain}{\mathcal{D}}
\newcommand{\vort}{\omega}
\newcommand{\bvort}{\boldsymbol{\omega}}
\newcommand{\leviciv}{\epsilon}
\newcommand{\partfun}{\mathcal{Z}}

\newcommand\Ro{\mbox{\textit{Ro}}}  
\newcommand\Fr{\mbox{\textit{Fr}}}  

\title{Restricted Equilibrium and the Energy Cascade in Rotating and Stratified Flows}

\author[C. Herbert, A. Pouquet and R. Marino]%
{Corentin Herbert$^1$%
  \thanks{Email address for correspondence: cherbert@ucar.edu},\ns
Annick Pouquet$^{1,2}$\break
and Raffaele Marino$^1$}

\affiliation{$^1$National Center for Atmospheric Research, P.O. Box 3000, Boulder, CO 80307, USA\\[\affilskip]
$^2$Department of Applied Mathematics, University of Colorado, Boulder, CO, 80309, USA}

\pubyear{2013}
\volume{}
\pagerange{}
\date{?; revised ?; accepted ?. - To be entered by editorial office}

\begin{document}

\maketitle

\begin{abstract}
Most of the turbulent flows appearing in nature (e.g. geophysical and astrophysical flows) are subjected to strong rotation and stratification. These effects break the symmetries of classical, homogenous isotropic turbulence. In doing so, they introduce a natural decomposition of phase space in terms of wave modes and potential vorticity modes. The appearance of a new time scale associated to the propagation of waves, in addition to the eddy turnover time, increases the complexity of the energy transfers between the various scales; nonlinearly interacting waves may dominate at some scales while balanced motion may prevail at others. In the end, it is difficult to predict \emph{a priori} if the energy cascades downscale as in homogeneous isotropic turbulence, upscale as expected from balanced dynamics, or follows yet another phenomenology.

In this paper, we suggest a theoretical approach based on equilibrium statistical mechanics for the ideal system, inspired from the restricted partition function formalism introduced in metastability studies. In this framework, we show analytically that in the presence of rotation, when the dynamics is restricted to the slow modes, the equilibrium energy spectrum features an infrared divergence characteristic of an inverse cascade regime, whereas this is not the case for purely stratified flows.
\end{abstract}



\section{Introduction}

Three dimensional homogeneous and isotropic turbulence (3D HIT) can be seen as essentially a competition between inertia (nonlinearity) and viscosity. The former is responsible for coupling all the different scales of motion, while the latter acts as a sink of energy at small scales. The standard phenomenology, known since~\cite{RichardsonBook} and~\cite{Kolmogorov1941a}, is that inertia acts essentially locally in an \emph{inertial range}, breaks big whorls to form smaller and smaller whorls in a process referred to as the direct cascade of energy, ultimately feeding molecular viscosity.
In many situations encountered in reality, forces other than inertia and viscosity may also be at work. For instance, a conducting fluid in a magnetic field will be subjected to the Lorenz force, the Coriolis force exerts itself on fluids subject to rotation, while buoyancy acts upon stratified fluids. These effects play a major part in astrophysical flows, while the latter two are essential ingredients of geophysical flows, like the atmosphere and the ocean. This study is devoted to this last case.

The presence of the Coriolis and buoyancy forces have several important consequences. First of all, they introduce linear terms in the equations of motion, which allow for the propagation of waves, respectively inertial waves and internal gravity waves, or when both are present at the same time, the more general family of inertia-gravity waves.
Besides, they also allow for regimes where the pressure force balances the new forces: geostrophic balance in the horizontal direction, when the pressure gradient balances the Coriolis force, and hydrostatic balance in the vertical direction, when the pressure gradient balances the buoyancy force. These balanced motions play a fundamental role in the theory of geophysical fluid dynamics~\citep{PedloskyGFD,SalmonBook,VallisBook,McWilliamsBook}, as the atmosphere and the oceans of the Earth are close to geostrophic and hydrostatic balance.

It has been suggested early on that balanced modes could be seen as a submanifold of phase space, consisting of the slow motions, hence the name \emph{slow manifold}~\citep{Lorenz1980,Leith1980}, and long standing questions have	 been whether this manifold is invariant --- an orbit initialized at a point on the manifold remains on the manifold --- and whether it is stable --- orbits initialized in the vicinity of the manifold converge to the manifold --- with the hope that it may coincide with the attractor of realistic models of geophysical fluid dynamics, like primitive equations~\citep{Lorenz1986,Lorenz1992,Warn1997}. Although these questions are difficult to answer mathematically, it seems that the spontaneous departures --- i.e. other than due to external wave generation mechanisms, like topography --- from the manifold are relatively small and infrequent~\citep{Vanneste2013}, at least when rotation and stratification are strong enough. This issue is of primary importance for numerical weather prediction, where gravity waves were initially referred to as noise, and two main approaches have been followed: either writing new dynamical equations which enforce in particular the balance relations at all times (\emph{quasi-geostrophic dynamics}) or initializing the system in a state devoid of such oscillations, on the \emph{slow manifold}~\citep{Baer1977,Machenhauer1977,Leith1980,Vautard1986}.
Recently, rigorous mathematical methods have been developed to derive families of balanced dynamics on such \emph{slow manifolds} through multiscale asymptotic expansions~\citep{Embid1998,Julien2006,Wingate2011}, thereby generalizing the quasi-geostrophic dynamics, which correspond to an asymptotic regime of low Rossby and Froude number in a small aspect ratio domain.

In general, these various types of motions coexist, and one may prevail in a given range of scales while the other does at other scales.
While geostrophic balance seems to be relatively robust at large scales, it may break down at smaller scales (e.g. the submesoscale in the ocean, which are becoming more and more studied as numerical models and observations permit). Wave dynamics may then drive the processes which occur at these scales, like vertical mixing in the ocean. In this range of scales, theoretical investigations have focused on the methods of weak turbulence (see the reference works by \cite{NazarenkoBook} and \cite{Newell2011} and references therein), which assumes that the amplitudes of the waves are slowly modulated by nonlinear interactions, with a mitigated agreement with observations~\citep{Polzin2011}. At yet smaller scales, it may be expected that isotropy should recover~\citep{Dubrulle1992,Zeman1994}, and the phenomenology of the turbulent transfers should comply with the standard Kolmogorov theory. In rotating turbulence, high resolution numerical simulations indeed support this view~\citep{Mininni2012}.
To these different ranges may correspond different energy spectrum power-laws, and even cascades with different directions.
In particular, it remains to be fully clarified under which circumstances, and through which mechanisms, rotation and/or stratification may lead to an inverse cascade.
While it seems relatively unambiguous from numerical simulations that rotating turbulence supports an inverse cascade~\citep{LSmith1996,LSmith1999b,Chen2005,Mininni2010a}, the picture is slightly more fuzzy in the presence of stratification. In purely stratified flows, a forward energy cascade has been observed in most numerical simulations~\citep{Godeferd1994,Waite2004,Lindborg2006,Brethouwer2007,Lindborg2007b,Marino2013b}, although a weak inverse cascade has been reported for strong stratification with a 2D forcing~\citep{Herring1989}. When rotation is added, an inverse cascade appears~\citep{Metais1996,LSmith2002,Marino2013b}, which is particularly strong when rotation and stratification are of comparable strengths. This inverse cascade of energy can coexist with a second inertial range corresponding to a downscale energy cascade~\citep{Pouquet2013}.

The purpose of this study is to provide some theoretical insight on the role of the slow modes in the emergence of an inverse cascade in rotating and/or stratified turbulence. To do so, we make use of a theoretical tool which has proved useful in studying the direction of the cascade in the case of homogeneous isotropic flows: equilibrium statistical mechanics.
In the case of 2D turbulence, \cite{Kraichnan1967,Kraichnan1975} (see also~\cite{Kraichnan1980}) introduced the canonical probability distribution for Galerkin truncated flows, based on conservation of two quadratic quantities: the energy and enstrophy. At statistical equilibrium, he identified a regime of high energy (relative to the enstrophy), in which a condensation of energy in the gravest modes was expected due to the appearance of negative temperatures. Comparing the slope of the energy and enstrophy spectra at equilibrium with that obtained by self-similarity requirements in two hypothetical inertial ranges, assuming a tendency for the flow to evolve towards equilibrium, even in the presence of forcing and dissipation, hints at the direction of the cascades in the inertial ranges: downscale cascade of enstrophy and upscale cascade of energy. This behavior has been largely confirmed since then (see e.g.~\cite{Boffetta2012} for a review). For 3D turbulence, the same methods yield an energy equipartition spectrum at statistical equilibrium~\citep{Lee1952,Kraichnan1973}, indicative of a forward cascade regime, even in the presence of helicity~\citep{Chen2003a}. The same methods were applied to geophysical flows (see~\cite{Lucarini2013} for a general review), in the context of topographic turbulence~\citep{Herring1977,Merryfield1996} and quasi-geostrophic flows with discrete or continuous stratification~\citep{Salmon1976, Holloway1986,Merryfield1998}, and also predict an inverse cascade in the horizontal direction, together with a tendency towards \emph{barotropization} of the flow.

The methods of absolute equilibrium for Galerkin truncated flows have been used beyond the framework of 2D and quasi-geostrophic turbulence. \cite{Errico1984} has adapted the methods to a primitive equations system, \cite{Warn1986} has considered a shallow-water model and \cite{Bartello1995} and \cite{Waite2004} turned to the Boussinesq equations. 
In all these cases, the average spectrum of energy at absolute equilibrium points at a forward cascade of energy, unlike in quasi-geostrophic dynamics. Besides, at absolute equilibrium, the energy tends to concentrate in the wave modes.
These results have been mostly interpreted in the context of the question of stability of the slow manifold and the connection with initialization methods has been underlined.

In this paper, in the context of Boussinesq flows, we provide a study of the statistics stemming from the nonlinear interactions taking into account either all the modes or only the slow modes. Our main result is that taking into account only the slow modes can lead to an inverse cascade for rotating and rotating/stratified turbulence, but not for stratified turbulence.
We suggest an analogy with the theory of metastability: in all rigor, the macroscopic behavior of the system is dominated by the contributions to the statistics of the microstates concentrating near the most probable macrostate. When metastable states exist, although the macroscopic system may remain for very long times in a state which is not the most probable macrostate, the corresponding regions of phase space contribute only subdominant contributions to the equilibrium statistics. Therefore, it was suggested by \cite{Penrose1971,Penrose1979} to consider the statistics resulting from constructing the equilibrium probability density restricted to a submanifold of phase space, containing the microstates which contribute to the metastable state but not those which concentrate around the absolute equilibrium state. This procedure has been applied to classical models of condensed matter physics, like the Ising model of ferromagnetism~\citep{Capocaccia1974} or the van der Waals-Maxwell model of the liquid-vapor transition~\citep{Penrose1971}, and more recently, to the case of turbulent flows, in the context of helically constrained flows~\citep{Herbert2013d}. Here, we adapt this idea to the \emph{slow manifolds} of rotating and/or stratified flows.

In a first step (\cref{slowmansec}), we introduce the slow and fast modes of rotating and/or stratified flows by finding the normal modes of the linearized equations, following \cite{Bartello1995}. We use this decomposition to define the \emph{slow manifold}, for each case. Then (\cref{mecastatsec}), we investigate the equilibrium statistics obtained by restricting the partition function to this submanifold. A question of major interest is: in which cases do the \emph{slow modes} yield an inverse (resp. direct) cascade of energy, and to what extent is this cascade similar to 2D (resp. 3D) turbulence?
Finally (\cref{discussec}), we discuss the range of validity of the hypotheses underlying the computations made in this paper.

\section{The Boussinesq equations and the slow manifold}\label{slowmansec}

\subsection{The Ideal Boussinesq Equations}\label{bousssec}

We will consider inviscid, incompressible flows in a tri-periodic cubic (with length $L$ --- unless mentioned otherwise we shall use $L=2\pi$) domain $\domain$ --- i.e. the torus $\mathbb{T}^3$ --- rotating around the axis $\vec{e_\parallel}$ with angular velocity $\Omega$, with an imposed stratification along the same axis. The flow is described by the inviscid Boussinesq equations:
\begin{align}
\partial_t \vec{u} + \vec{u} \cdot \bnabla \vec{u} &= -\bnabla P -2 \vec{\Omega} \times \vec{u} -N\theta \vec{e_\parallel},\label{bousseq1}\\
\partial_t \theta + \vec{u} \cdot \bnabla \theta &= N u_\parallel,\label{bousseq2}\\
\bnabla \cdot \vec{u} &=0,\label{bousseq3}
\end{align}
where $\vec{u}$ is the velocity field, $u_\parallel=\vec{u} \cdot \vec{e_\parallel}$ its projection on the direction of rotation, $P$ the pressure field, $\theta$ the buoyancy, and $N=\sqrt{-g\partial_z \bar{\theta}/\theta}$ the Brunt-V\"ais\"al\"a frequency, with $\bar{\theta}$ the imposed stratification profile. Note that we are not considering any buoyancy diffusion. We also define the Coriolis frequency $f=2\Omega$.

All along this paper, we shall consider three cases: the full Boussinesq equations \eqref{bousseq1}-\eqref{bousseq3} describing rotating and stratified flows ($N \neq 0$, $f\neq 0$), but also the particular case of purely stratified flows with no rotation ($f=0$, $N \neq 0$, which simply has the effect of suppressing the term $\vec{\Omega} \times \vec{u}$ in \eqref{bousseq1}) and of purely rotating flows with no imposed stratification ($N=0$, $f \neq 0$; in this case the buoyancy becomes a passive scalar, the behavior of which we shall not be interested in here).

As is customary, we introduce the Fourier decompositions for the various fields of interest:
\begin{align}
u_i (\vec{x}) 	&= \sum_{\vec{k} \in \wavevecset{B}} \fourier{u}_i(\vec{k}) e^{i \vec{k} \cdot \vec{x}},\\
\theta(\vec{x}) 	&= \sum_{\vec{k} \in \wavevecset{B}} \fourier{\theta}(\vec{k}) e^{i \vec{k} \cdot \vec{x}}.
\end{align}
Here, $\wavevecset{B}$ denotes the set of wave vectors, which is \emph{a priori} $2\pi/L \mathbb{Z}^3$ where $L$ is the length of the domain. We have implicitly introduced a basis $(\vec{e_1},\vec{e_2},\vec{e_3})$ of $\mathbb{R}^3$, so that $\vec{u}(\vec{x}) = u^i (\vec{x}) \vec{e_i}$, where we have adopted the Einstein convention of summation of repeated indices. For simplicity, we make the choice $\vec{e_3}=\vec{e_\parallel}$.
The dynamical equations can be recast in Fourier space:
\begin{align}
\partial_t \fourier{u}_i(\vec{k}) &= -\frac{i}{2} \mathcal{P}_i^{\phantom{i}jl}(\vec{k}) \sum_{\vec{p}+\vec{q}=\vec{k}} \fourier{u}_j(\vec{p}) \fourier{u}_l(\vec{q}) +f P_{ij}(\vec{k}) \leviciv^{jl3}\fourier{u}_l(\vec{k})-N P_{i3}(\vec{k}) \fourier{\theta}(\vec{k}),\\
\partial_t \fourier{\theta}(\vec{k}) &=-i \sum_{\vec{p}+\vec{q}=\vec{k}}  k^j \fourier{u}_j(\vec{p})\fourier{\theta}(\vec{q})+N \fourier{u}_3(\vec{k}),\\
k^i \fourier{u}_i(\vec{k}) &=0,
\end{align}
where the projection operator is given by $\mathcal{P}_{ijl}(\vec{k})=k_j P_{il}(\vec{k})+k_l P_{ij}(\vec{k})$, with $P_{ij}(\vec{k})=\delta_{ij}-k_ik_j/k^2$, and $\leviciv$ is the standard, rank 3, totally asymmetric, Levi-Civita tensor.

As usual for the inviscid system, the energy of the flow is conserved:
\begin{align}
E &= \frac 1 2 \int_\domain ( \| \vec{u}(\vec{x}) \|^2 + \theta(\vec{x})^2)d\vec{x},\\
&= \frac 1 2 \sum_{\vec{k} \in \wavevecset{B}} ( \fourier{u}_i(\vec{k}) \fourier{u}^i(\vec{k})^* +\abs{\fourier{\theta}(\vec{k})}^2).\label{fourierenergyeq}
\end{align}
In addition, we can introduce the potential vorticity
\begin{align}
q &= f \partial_\parallel \theta - N \vort_\parallel + \bvort \cdot \bnabla \theta,
\end{align}
which is conserved along Lagrangian trajectories:
\begin{align}
\partial_t q + \vec{u} \cdot \bnabla q &= 0.
\end{align}
As a consequence, the integral over the whole domain of any function of potential vorticity is a global invariant: for a function $g$, let $I_g = \int_\domain g(q(\vec{x}))d\vec{x}$, then $\dot{I_g}=0$.
In particular, all the moments of the potential vorticity fields are global invariants corresponding to $g(x)=x^n$. This is analogous to the case of two-dimensional turbulence and also quasi-geostrophic turbulence. In both these cases, the particular case $n=2$ yields a second quadratic invariant, which plays a major part in understanding the phenomenology of the nonlinear transfers of energy~\citep{Kraichnan1967,Rhines1979,Kraichnan1980,SalmonBook}. Here, the case $n=2$ does not lead to a quadratic quantity because of the nonlinear term $\bvort \cdot \bnabla \theta$ in the definition of potential vorticity. 
In Fourier space, this corresponds to a convolution term in the Fourier coefficients of potential vorticity:
\begin{align}
\fourier{q}(\vec{k}) &= if k_\parallel \fourier{\theta}(\vec{k})-iN \leviciv_{jlm}k^j \fourier{u}^l(\vec{k})\delta_3^m-\sum_{\vec{p}+\vec{q}=\vec{k}} \leviciv_{jlm} q^j p^l \fourier{u}^m(\vec{p}) \fourier{\theta}(\vec{q}).
\end{align}
Neglecting the nonlinear contribution to potential vorticity (see section \ref{pvdiscussec}), the potential enstrophy reduces to a quadratic quantity:
\begin{align}
\Gamma_2 &= \frac 1 2 \int_\domain (f\partial_\parallel \theta-N\vort_\parallel)^2, \label{potenstrophydefeq}\\
&= \frac 1 2 \sum_{\vec{k} \in \wavevecset{B}} \abs{f k_\parallel \fourier{\theta}(\vec{k})-N \leviciv_{jlm}k^j \fourier{u}^l(\vec{k})\delta_3^m}^2,\\
&= \frac 1 2 \sum_{\vec{k} \in \wavevecset{B}} \Big\lbrack f^2 k_\parallel^2 \abs{\fourier{\theta}(\vec{k})}^2 + N^2 \leviciv_{jl3}\leviciv_{pm3} k^j k^p \fourier{u}^l(\vec{k})\fourier{u}^m(\vec{k})^*-fNk_\parallel \leviciv_{jl3}k^j [\fourier{u}^l(\vec{k})\fourier{\theta}(\vec{k})^*+ \fourier{u}^l(\vec{k})^*\fourier{\theta}(\vec{k})] \Big\rbrack.\label{fourierenstrophyeq}
\end{align}

From \cref{fourierenergyeq} and \cref{fourierenstrophyeq}, we see that the generic form for the quadratic global invariant in that case is
\begin{align}
\beta E + \alpha \Gamma_2 &= \frac 1 2 \sum_{\vec{k} \in \wavevecset{B}} {^t\vec{X}(\vec{k})^*} \vec{M}_{\beta,\alpha}(\vec{k}) \vec{X}(\vec{k}),
\end{align}
for arbitrary $\alpha,\beta \in \mathbb{R}$, and with
\begin{align}
\vec{X}(\vec{k}) &= \begin{pmatrix} \fourier{u}_1(\vec{k}) \\ \fourier{u}_2(\vec{k}) \\ \fourier{u}_3(\vec{k}) \\ \fourier{\theta}(\vec{k}) \end{pmatrix},
& \vec{M}_{\beta,\alpha}(\vec{k}) &= 
\begin{pmatrix}  
\beta+\alpha N^2 k_2^2 	& -\alpha N^2 k_1 k_2 		& 0 		& \alpha f N k_2 k_3 \\
-\alpha N^2 k_1 k_2 		& \beta+\alpha N^2 k_1^2 	& 0		& -\alpha f N k_1 k_3 \\
0					& 0						& \beta	& 0\\
\alpha f N k_2 k_3		& -\alpha f N k_1 k_3			& 0		& \beta+\alpha f^2 k_3^2
\end{pmatrix}.\label{invmatrixeq}
\end{align}

\subsection{Slow manifold: wave and vortical modes}\label{phasespacesec}

\subsubsection{Rotating and stratified case ($N \neq 0$ and $f \neq 0$)}\label{stratrotphasespacesec}

The ideal Boussinesq equations described in \cref{bousssec} can be seen as a dynamical system evolving on a phase space $\Lambda$. In real space, $\Lambda$ would be the product of the manifold containing all the incompressible velocity fields on the flat torus with finite energy and of the space of buoyancy fields on the flat torus with finite energy: $\Lambda= \mathcal{M}\times L^2(\domain)$, where $\mathcal{M} = \{ \vec{u} \in L^2(\domain), \nabla \cdot \vec{u} = 0\}$. 
In Fourier space, $\Lambda$ can be defined in a dual manner as the product of the Fourier amplitude \enquote{fields} on the Pontryagin dual of the flat torus $\pdual{\mathcal{T}}$ for both the velocity and the buoyancy. One has to incorporate the reality condition $\forall \vec{k} \in \wavevecset{B}, \fourier{u}_i(\vec{k})=\fourier{u}_i(-\vec{k})^*, \fourier{\theta}(\vec{k})=\fourier{\theta}(-\vec{k})^*$ for both fields, and the incompressibility condition $\forall \vec{k} \in \wavevecset{B}, \fourier{u}_i(\vec{k})k^i=0$ for the velocity field: 
\begin{equation}
\Lambda = \{ (\vec{\fourier{u}},\fourier{\theta}) \in L^2(\pdual{\mathcal{T}})\times L^2(\pdual{\mathcal{T}}), \forall \vec{k} \in \pdual{\mathcal{T}}, \fourier{u}_i(\vec{k})=\fourier{u}_i(-\vec{k})^*, k^i\fourier{u}_i(\vec{k})=0 \text{ and } \fourier{\theta}(\vec{k})=\fourier{\theta}(-\vec{k})^*\}.
\end{equation}

The purpose of this section is to provide a description of phase space which takes into account the fact that some modes have a different physical nature than other modes. To achieve this goal, we follow the method of \cite{Leith1980}, adapted to the Boussinesq equations by \cite{Bartello1995}, which consists in finding the normal modes of the tangent map. The linearized evolution equations read:
\begin{align}
\partial_t \fourier{u}_i(\vec{k}) &= f P_{ij}(\vec{k}) \leviciv^{jl3}\fourier{u}_l(\vec{k})-N P_{i3}(\vec{k}) \fourier{\theta}(\vec{k}),\label{lindyneq1}\\
\partial_t \fourier{\theta}(\vec{k}) &=N \fourier{u}_3(\vec{k}),\label{lindyneq2}
\end{align}
or in matrix form
\begin{align}
\dot{\vec{X}}(\vec{k}) &= \vec{L}(\vec{k})\vec{X}(\vec{k}),
\intertext{with $\vec{X}(\vec{k})$ defined by \eqref{invmatrixeq} and}
\vec{L}(\vec{k})&=\begin{pmatrix}
f \frac{k_1k_2}{k^2}			& f\Big( 1-\frac{k_1^2}{k^2}\Big)		& 0		& N \frac{k_1k_3}{k^2} \\
-f\Big( 1-\frac{k_2^2}{k^2}\Big)	& -f \frac{k_1k_2}{k^2}			& 0		& N \frac{k_2k_3}{k^2} \\
f \frac{k_2k_3}{k^2}			& -f \frac{k_1k_3}{k^2}			& 0		& -N \Big( 1- \frac{k_3^2}{k^2}\Big) \\
0						& 0							& N		& 0
\end{pmatrix}.
\end{align}
To get rid of the incompressibility condition, it is convenient to introduce new variables: following \cite{Bartello1995}, we use the horizontal divergence and vertical component of vorticity:
\begin{align}
\delta(\vec{k}) &= ik_1 \fourier{u}_1(\vec{k})+ik_2\fourier{u}_2(\vec{k})=-ik_\parallel\fourier{u}_3(\vec{k}),\\
\zeta(\vec{k}) &= i k_1 \fourier{u}_2(\vec{k}) - i k_2 \fourier{u}_1(\vec{k})=\fourier{\vort}_3(\vec{k}),
\intertext{together with the rescaled variables}
D(\vec{k}) &= \frac k {k_\parallel} \delta(\vec{k}) =-i k \fourier{u}_3(\vec{k}),\\
T(\vec{k}) &= -k_\perp \fourier{\theta}(\vec{k}).
\end{align}
Now, the linear evolution reads
\begin{align}
\dot{\vec{Z}}(\vec{k}) &= \vec{L''}(\vec{k})\vec{Z}(\vec{k}),&
\intertext{with}
\vec{Z}(\vec{k})&=\begin{pmatrix} \zeta(\vec{k}) \\ D(\vec{k}) \\ T(\vec{k}) \end{pmatrix},& \vec{L''}(\vec{k})=\begin{pmatrix}
0				& -f \frac{k_\parallel}{k}	& 0	 				\\
f \frac{k_\parallel}{k}	& 0					& -i N \frac{k_\perp}{k} 	\\
0				& -i N \frac{k_\perp}{k}	& 0		 			\\
\end{pmatrix}.
\end{align}
The spectrum of the matrix $\vec{L''}(\vec{k})$ is $\Sp \vec{L''}(\vec{k}) = \{0, i \sigma(\vec{k}), -i \sigma(\vec{k})\}$, where 
\begin{align}
\sigma(\vec{k})&= \sqrt{f^2 \frac{k_\parallel^2}{k^2}+N^2 \frac{k_\perp^2}{k^2}} >0
\end{align}
is the inertia-gravity waves frequency. In particular, it is a singular matrix.
Note that the matrix $\vec{L''}(\vec{k})$ is anti-hermitian: $^t \vec{L''}(\vec{k})^*=-\vec{L''}(\vec{k})$ ($i\vec{L''}(\vec{k})$ is hermitian). In particular, it is a normal matrix, and therefore it is unitarily similar to a diagonal matrix: there exists a unitary matrix $\vec{P} \in U(3)$ such that 
\begin{align}
\vec{L''}(\vec{k}) &= \vec{P} \begin{pmatrix} -i \sigma(\vec{k}) & 0 & 0 \\ 0 & 0 & 0 \\ 0 & 0 & i\sigma(\vec{k}) \end{pmatrix} \vec{P}^{-1}.
\end{align}
We introduce the eigenvectors $\vec{Z_0}(\vec{k})$ corresponding to the eigenvalue $0$, and $\vec{Z_\pm}(\vec{k})$ corresponding to the eigenvalue $\pm i \sigma(\vec{k})$: $\vec{L''}(\vec{k}) \vec{Z_0}(\vec{k})=0$, $\vec{L''}(\vec{k}) \vec{Z_\pm}(\vec{k})=\pm i \sigma(\vec{k})$.
In the canonical basis, these (normalized) eigenvectors are given by
\begin{align}
\vec{Z_0}(\vec{k}) &= \frac{1}{k \sigma(\vec{k})} \begin{pmatrix} iN k_\perp \\ 0\\ fk_\parallel \end{pmatrix}, & \vec{Z_\pm}(\vec{k}) &= \frac{1}{\sqrt{2} k \sigma(\vec{k})} \begin{pmatrix} -if k_\parallel \\ \mp k \sigma(\vec{k}) \\ N k_\perp\end{pmatrix}.
\end{align}
These eigenvectors form an orthonormal basis of $\mathbb{C}^3$, and we can decompose every vector as $\vec{Z}(\vec{k})= a_0(\vec{k}) \vec{Z_0}(\vec{k}) + a_-(\vec{k}) \vec{Z_-}(\vec{k}) + a_+(\vec{k}) \vec{Z_+}(\vec{k})$. 

Physically speaking, this means that the linear terms of the equation of motion leave the mode $\vec{Z_0}(\vec{k})$ unchanged: $\dot{a}_0(\vec{k})=0$; of course, this does not remain true when considering the full equations (see section \ref{slowdiscussec}). On the contrary, the modes $\vec{Z_\pm}(\vec{k})$ correspond to waves propagating with frequency $\pm \sigma(\vec{k})$. The nonlinear dynamics is going to couple the evolution of these slow and fast modes in a complicated way.
Nevertheless, they provide insight on the structure of phase space, and one may define submanifolds of phase space corresponding to retaining only the slow or fast modes: we denote $\Lambda_0$ the submanifold of $\Lambda$ defined by the conditions $a_+(\vec{k})=a_-(\vec{k})=0$, and $\Lambda_W$ the submanifold defined by the condition $a_0(\vec{k})=0$. Clearly, $\Lambda=\Lambda_0 \oplus \Lambda_W$.

Note that the slow manifold corresponds to balanced motion. Indeed, writing the linearized dynamics in Fourier space (Eqs. \ref{lindyneq1}-\ref{lindyneq2}) including explicitly the pressure term --- instead of using the projection operator --- we obtain the relation between pressure and the dynamical variables which is satisfied by the linear dynamics:
\begin{align}
\fourier{P}(\vec{k}) &= - \frac{f \zeta(\vec{k})+ i N k_\parallel T(\vec{k})/k_\perp}{k^2}.
\intertext{The slow modes satisfy the relation $iN k_\perp T(\vec{k})=f k_\parallel \zeta(\vec{k})$, which imply that the Fourier coefficients of pressure can be rewritten in the two equivalent forms:}
\fourier{P}(\vec{k}) &= - \frac{f\zeta(\vec{k})}{k_\perp^2},\\
&= -i\frac{NT(\vec{k})}{k_\parallel k_\perp}.
\end{align}
The first of these two forms correspond to geostrophic balance, and the second to hydrostatic balance. Therefore, the (linear) dynamics on the slow manifold is both geostrophically and hydrostatically balanced. The true, nonlinear, dynamics can of course destroy these balance relations.

Mathematical details about the change of variable described above are given in \Cref{normalmodessec}. Note that the cases $k_\parallel=0$ and $k_\perp=0$ may seem problematic. When $k_\parallel=0$, $\delta(\vec{k})=0$, but the appropriate dynamical variable is then $D(\vec{k})$, which is proportional to the vertical velocity. By contrast, when $k_\perp=0$ (which implies $k_1=k_2=0$), the change of variables breaks down: $\delta(\vec{k})=\zeta(\vec{k})=D(\vec{k})=T(\vec{k})=0$, and so does the alternative description proposed in \Cref{normalmodessec}. For such modes, the incompressibility condition imposes that the vertical velocity vanishes: $\fourier{u}_3(\vec{k})=0$. Therefore, the linearized dynamics keep $\fourier{\theta}(\vec{k})$ constant. The two horizontal components of velocity $\fourier{u}_1(\vec{k}),\fourier{u}_2(\vec{k})$ oscillate with frequency $f$. In this situation, the flow looks like vertically sheared copies of horizontally uniform 2D flows, hence they are often referred to as \emph{shear} modes or \emph{pancake} modes. These modes are hydrostatically balanced. Their slow component is trivial: the velocity field vanishes completely. Nevertheless, these modes are coupled to the others in the full nonlinear system. Their modal contribution to the energy and to the potential enstrophy are respectively $(\abs{\fourier{u}_1(\vec{k})}^2+\abs{\fourier{u}_2(\vec{k})}^2+\abs{\fourier{\theta}(\vec{k})}^2)/2$ and $f^2 k_\parallel^2 \abs{\fourier{\theta}(\vec{k})}^2/2$. We introduce the set of wavevectors corresponding to the shear modes: $\wavevecset{B}_S = \{ \vec{k} \in \wavevecset{B}, k_\perp=0 \}$.

The change of basis introduced here also diagonalize simultaneously the quadratic invariants of the system.
Indeed,
\begin{align}
E&=\frac 1 2 \sum_{\vec{k} \in \wavevecset{B}} {^t \vec{X}(\vec{k})^*} \vec{X}(\vec{k}),\\
&= \frac 1 2 \sum_{\vec{k} \in \wavevecset{B}_S^c} \frac{{^t \vec{Z}(\vec{k})^*} \vec{Z}(\vec{k})}{k_\perp^2}+\frac 1 2 \sum_{\vec{k} \in \wavevecset{B}_S} \frac{\abs{\fourier{u}_1(\vec{k})}^2+\abs{\fourier{u}_2(\vec{k})}^2+\abs{\fourier{\theta}(\vec{k})}^2}{2},\\
&=\frac 1 2 \sum_{\vec{k} \in \wavevecset{B}_S^c} \frac{(\abs{a_0(\vec{k})}^2 +\abs{a_+(\vec{k})}^2+\abs{a_-(\vec{k})}^2)}{k_\perp^2} + \frac 1 2 \sum_{\vec{k} \in \wavevecset{B}_S} \frac{\abs{\fourier{u}_1(\vec{k})}^2+\abs{\fourier{u}_2(\vec{k})}^2+\abs{\fourier{\theta}(\vec{k})}^2}{2},\\
&=\frac 1 2 \sum_{\vec{k} \in \wavevecset{B}} (\abs{A_0(\vec{k})}^2 +\abs{A_+(\vec{k})}^2+\abs{A_-(\vec{k})}^2),
\intertext{with}
A_0(\vec{k})&=\begin{cases}
\fourier{\theta}(\vec{k}) & \text{if } \vec{k} \in \wavevecset{B}_S \\
\frac{a_0(\vec{k})}{k_\perp} & \text{if } \vec{k} \in \wavevecset{B}_S^c
\end{cases},
\qquad
A_\pm(\vec{k})=\begin{cases}
(\fourier{u}_2(\vec{k}) \pm i \fourier{u}_1(\vec{k}))/\sqrt{2} & \text{if } \vec{k} \in \wavevecset{B}_S \\
\frac{a_\pm(\vec{k})}{k_\perp} & \text{if } \vec{k} \in \wavevecset{B}_S^c
\end{cases},
\intertext{and where $X^c$ denotes the complement of the set $X$. Similarly,}
\Gamma_2 &= \frac 1 2 \sum_{\vec{k} \in \wavevecset{B}} k^2 \sigma(\vec{k})^2 \abs{A_0(\vec{k})}^2,
\intertext{so that}
\beta E + \alpha \Gamma_2 &= \frac 1 2 \sum_{\vec{k} \in \wavevecset{B}} {^t\vec{A}(\vec{k})^*} \vec{\Delta}_{\beta,\alpha}(\vec{k}) \vec{A}(\vec{k}),
\intertext{with}
\vec{\Delta}_{\beta,\alpha}(\vec{k}) &= \begin{pmatrix}
\beta & 0 & 0 \\
0 & \beta+\alpha k^2 \sigma(\vec{k})^2 & 0 \\
0 & 0 & \beta
\end{pmatrix},\qquad  \vec{A}(\vec{k}) = \begin{pmatrix} A_-(\vec{k}) \\ A_0(\vec{k}) \\ A_+(\vec{k}) \end{pmatrix}.
\end{align}
Note that only the slow modes contribute to potential enstrophy. Therefore, they should also be referred to as \emph{vortical modes}, by opposition to the \emph{wave modes}, which provide a vanishing contribution to the potential enstrophy, no matter how much energy they contain.

\subsubsection{Purely stratified case ($N \neq 0$ and $f=0$)}

The analysis of \cref{stratrotphasespacesec} remains valid in the case of purely stratified flows, and it suffices to set $f=0$ in the above equations. In particular, the normal modes of the linearized dynamics still consist of a \emph{slow mode} $\vec{Z_0}(\vec{k})$, now given directly by the vertical component of vorticity $\zeta(\vec{k})$, and two gravity wave modes $\vec{Z_\pm}(\vec{k})$, with a dispersion relation $\sigma(\vec{k})=N k_\perp/k$:
\begin{align}
\vec{Z_0}(\vec{k}) &= \begin{pmatrix} 1 \\ 0\\ 0 \end{pmatrix}, & \vec{Z_\pm}(\vec{k}) &= \frac{1}{\sqrt{2}} \begin{pmatrix} 0 \\ \mp 1 \\ 1 \end{pmatrix}.
\end{align}
As in the case with rotation, we shall denote by $\Lambda_0$ the \emph{slow manifold}, and by $\Lambda_W$ the wave manifold.

A difference with the rotating-stratified case is that now, the slow modes do not satisfy the hydrostatic balance equations in general, unless $k_\perp=0$. In other words, the shear modes coincide with the hydrostatically balanced modes, and they have a trivial linear dynamics; they are constant in time.

The quadratic invariants of course keep the same diagonal form as above. In particular, only the slow modes contribute to the potential enstrophy, and therefore will also be referred to as \emph{vortical modes}. Note, however, that the contribution of the shear modes to potential vorticity now vanishes.

\subsubsection{Purely rotating case ($N=0$ and $f \neq 0$)}

As mentioned above, we shall not be interested here in the behavior of the scalar field, which becomes passive when $N=0$. Hence in this case the full phase space is simply $\Lambda=\mathcal{M}$, with the above notations.
The above analysis of the linearized equations carries over to this case by simply getting rid of the $\fourier{\theta}$ or $T$ variable, and setting $N=0$. Therefore, the linear dynamics reads
\begin{align}
\dot{\vec{Z}}(\vec{k}) &= \vec{L''}(\vec{k})\vec{Z}(\vec{k}),&
\intertext{with}
\vec{Z}(\vec{k})&=\begin{pmatrix} \zeta(\vec{k}) \\ D(\vec{k}) \end{pmatrix},& \vec{L''}(\vec{k})=\begin{pmatrix}
0				& -f \frac{k_\parallel}{k}	\\
f \frac{k_\parallel}{k}	& 0					\\
\end{pmatrix}.
\end{align}
This time the spectrum of the matrix $\vec{L''}(\vec{k})$ is $\pm i \sigma(\vec{k})= \pm i f k_\parallel/k$. This is the dispersion relation of inertial waves. By contrast to the rotating-stratified case, the \emph{slow modes} do not exist for arbitrary $\vec{k}$, but only for $k_\parallel=0$. Therefore, the modes which are going to play here the role analogous to the vortical modes in rotating-stratified turbulence are the \emph{2D modes}. These are the only modes for which there is no wave propagation. We shall denote by $\Lambda_{2D}$ the submanifold of these slow modes, defined by $\fourier{u}_i(\vec{k})=0$ when $k_\parallel \neq0$. We further introduce the set of wavevectors corresponding to the 2D modes: $\wavevecset{B}_{2D} = \{ \vec{k} \in \wavevecset{B}, k_\parallel=0\}$.

The energy is expressed as previously:
\begin{align}
E &= \frac 1 2 \sum_{\vec{k} \in \wavevecset{B}_{2D}} \frac{{^t \vec{Z}(\vec{k})^*} \vec{Z}(\vec{k})}{k_\perp^2},\\
&= \frac 1 2 \sum_{\vec{k} \in \wavevecset{B}_{2D}} \frac{\abs{D(\vec{k})}^2 + \abs{\zeta(\vec{k})}^2 }{k_\perp^2}.
\end{align}
When $N=0$, the expression of $\Gamma_2$ given in \cref{potenstrophydefeq} does not impose any dynamical constraint on rotating flows. Nevertheless, the dynamics on the slow manifold $\Lambda_{2D}$ conserves $\int_\domain \vort_\parallel^2$. Indeed, on this manifold, the vertical velocity behaves as a passive scalar, while the horizontal part of the flow evolves as a 2D incompressible flow:
\begin{align}
\partial_t u_\parallel + \vec{u}_\perp \cdot \bnabla_\perp u_\parallel &=0\\
\partial_t \vec{u}_\perp + \vec{u}_\perp \cdot \bnabla_\perp \vec{u}_\perp &= -\bnabla_\perp P,\\
\bnabla_\perp \cdot \vec{u}_\perp &= 0.
\end{align}
It follows that the vertical component of vorticity is constant along streamlines: $\partial_t \vort_\parallel + \vec{u}_\perp \cdot \bnabla_\perp \vort_\parallel =0$, and thus the integral $\int_\domain g(\vort_\parallel)$ of any function $g$ of the vertical component of vorticity is a dynamical invariant. In particular, we shall consider here the quadratic invariant which plays the role of potential enstrophy in the previous cases:
\begin{align}
\Gamma_2^{2D} &= \frac 1 2 \int_\domain \vort_\parallel^2,\\
&= \frac 1 2 \sum_{\vec{k} \in \wavevecset{B}_{2D}} \abs{\zeta(\vec{k})}^2.
\end{align}
Besides, as the vertical component of velocity evolves as a passive scalar, the kinetic energy in the vertical and horizontal directions are conserved independently. Therefore, we have the two independent energy invariants:
\begin{align}
E_{2D} &= \frac 1 2 \sum_{\vec{k} \in \wavevecset{B}_{2D}} \frac{\abs{\zeta(\vec{k})}^2}{k_\perp^2} , & E_\parallel &= \frac 1 2 \sum_{\vec{k} \in \wavevecset{B}_{2D}} \frac{\abs{D(\vec{k})}^2}{k^2},
\end{align}
with $E=E_{2D}+E_\parallel$.

In principle there is an additional invariant: like in homogeneous isotropic flows, rotating flows conserve helicity~\citep{Montgomery1982}. Non vanishing helicity in rotating flows, which partly depletes the nonlinear interactions, has an impact on several aspects, such as small scale structures and intermittency~\citep{Mininni2010b,Pouquet2010}, and energy decay rate~\citep{Teitelbaum2009}. However, it does not have an impact on the direction of the energy cascade, as in the homogeneous isotropic case~\citep{Kraichnan1973}. Therefore, we shall not take it into account here. Note, however, that it would be possible to compute a partition function taking into account helicity conservation, but the simplest method to do so is to use the decomposition of the velocity field in terms of helical waves~\citep{Waleffe1992}, which is maybe not as well suited to separate the respective roles of the 2D modes and inertial waves at equilibrium.

The long time properties of such ideal flows have been examined numerically. The Coriolis force being linear does not impose a new constraint on statistical equilibria which must therefore be identical to the non-rotating case; in particular, they should not display any inverse cascade and large-scale condensation. However, at intermediate times $t_{**}$, the larger the stronger the rotation, inverse transfer clearly occurs in such flows, with a $t_{**} \sim \Omega^{3/4}$ dependency~\citep{Mininni2011a}.

\section{Restricted partition function}\label{mecastatsec}

\subsection{General formalism}

The fundamental idea of equilibrium statistical mechanics is to construct a probability density on the phase space of a dynamical system based only on the invariants of the dynamics. As soon as the Liouville theorem --- which states that the volume in phase space is conserved by the dynamics --- holds, such a probability measure is automatically an invariant measure. Among this class of invariant measures, different choices can be made for the dependence on the conserved quantities. Classical choices correspond to the standard \emph{ensembles} of statistical mechanics, and are solutions of the associated variational problems. In the microcanonical ensemble, the probability density is uniform on all the states with a given set of values for the invariants and vanishes for any other set of values. It is the appropriate distribution to describe an isolated system, which does not exchange with its surroundings. For instance, in the case of Boussinesq flows described above, taking into account only the quadratic invariants, the microcanonical density would read
\begin{equation}
\rho_{E^0,\Gamma_2^0} = \frac 1 {\Omega(E^0,\Gamma_2^0)} \delta(E-E^0) \delta(\Gamma_2-\Gamma_2^0),
\end{equation}
where the normalization factor $\Omega(E^0,\Gamma_2^0)$, referred to as the \emph{structure function}~\citep{KhinchinBook}, measures the volume of phase space occupied by microstates with energy $E^0$ and potential enstrophy $\Gamma_2^0$.

By contrast, the canonical ensemble allows for fluctuations of the invariant quantities (e.g. the energy), due to exchanges with the surroundings.
This description is relevant for systems which are in contact with a \emph{reservoir} (e.g. a \emph{thermostat}), which only fixes the average value of the invariant quantity, or equivalently, the conjugate Lagrange multiplier (e.g. the \emph{temperature}). For instance, for Boussinesq flows considered in this paper, taking into account only the quadratic invariants, the canonical density is given by
\begin{align}
\rho_{\beta,\alpha} &= \frac 1 {\partfun} e^{-\beta E - \alpha \Gamma_2},\label{canopdfdefeq}
\intertext{where the normalization factor $\partfun$, called the \emph{partition function} is given by}
\partfun &= \int_\Lambda e^{-\beta E - \alpha \Gamma_2} d\mu_\Lambda, \label{partfundefeq}
\end{align}
and encodes the statistics of the system.

For systems with long-range interactions, like hydrodynamic systems --- but also many others~\citep{DauxoisLRIbook,Campa2009} --- the microcanonical and canonical ensembles do not necessarily yield equivalent predictions~\citep{Kiessling1997,Ellis2000,Touchette2004}, even in the \emph{thermodynamic limit} --- i.e. for a large system. However, it is in general more difficult to work directly in the microcanonical ensemble, and until recently, applications of statistical mechanics to turbulent flows have focused on the canonical framework. In this paper, we are not going to delve further into this issue and will restrict ourselves to the canonical ensemble. For more information about the connection between the different statistical ensembles in the framework of turbulent flows, the reader is referred to~\cite{Bouchet2008,Venaille2011a,Herbert2012a,Herbert2012b,Herbert2013b,Bouchet2012}. 

Note also that in the case of turbulent flows, it may happen that other invariants exist, which may not be quadratic (this is the case of 2D turbulence, QG turbulence, but also shallow water flows or the Boussinesq equations considered here). Clearly, quadratic invariant make things simpler in that the canonical density is then Gaussian. Taking into account higher order invariants is possible, and several approaches have been suggested, in connection with the remark above on non-equivalence of the different statistical ensembles~\citep[e.g.][]{Miller1990,Robert1991a,Turkington1999}. For simplicity, we shall restrict here to the role of quadratic invariants.

As mentioned in the introduction, absolute equilibrium techniques and canonical probability distributions have been used in a variety of turbulent flows to study the direction of the cascade of energy in the forced-dissipative system. The prototypical cases are 2D and 3D HIT. In 2D turbulence, the partition function can be computed exactly. It is defined when the Lagrange parameters $(\beta,\alpha)$ belongs to a certain subset of $\mathbb{R}^2$, which can be decomposed in three subsets corresponding to different physical behaviors~\citep{Kraichnan1980}. Perhaps the most interesting case is that of \emph{negative temperature}: $\beta <0$, or equivalently, large energy compared to the enstrophy. In that case, the equilibrium energy spectrum features a divergence at large scales indicative of an inverse cascade of energy and a tendency towards spectral condensation. Note that in general, negative temperature states in the statistical physics of systems with long-range interactions are characteristic of organized states; one way to look at it is that the thermodynamic entropy becomes a decreasing function of energy: $\beta = \partial S/\partial E <0$. Therefore, the system can at the same time minimize its energy and maximize its entropy, in contrast to the usual energy-entropy competition which leads to the picture of order at small (positive) temperature, where the energy dominates the free energy, and disorder at large (positive) temperature, where the entropy dominates. On the contrary, 3D HIT does not exhibit such negative temperature states, even when the conservation of helicity is taken into account~\citep{Kraichnan1973}. The equilibrium energy spectrum, in that case, points at a forward cascade of energy. Therefore, the sign of the Lagrange parameters play a crucial role in the statistical equilibrium properties of the system, and by extension in the hypothetical cascade directions that they underlie.

The integral (\ref{partfundefeq}) defining the partition function is dominated by the contribution of the microstates which concentrate near the equilibrium state. In the presence of metastable states --- local minima of the free energy --- the contribution of the corresponding microstates to the partition function --- and hence, to the equilibrium statistics --- is subdominant. This does not reflect accurately the role that these metastable states can play in the dynamics, as they can potentially be very long lived. To study metastable states, it was suggested by~\cite{Penrose1971,Penrose1979} that one should restrict the integral to a subspace $\Lambda' \subset \Lambda$ of phase space excluding the microstates which concentrate near the equilibrium state. That way, the restricted partition function and the resulting restricted canonical probability density reflect the statistics that should be observed when the system is stuck in this metastable state. 

The application of equilibrium statistical mechanics methods to turbulent flows is peculiar in that most of the time, the equilibrium results on the ideal system are used to obtain insight on the behavior of the real, forced-dissipative system.
This means that the metastability interpretation does not carry over in a straightforward manner. However, metastable states can be replaced in a forced-dissipative framework by \emph{quasi-stationary states}. Besides, restricting the equilibrium procedure to a submanifold of phase space remains a valuable tool to determine the statistics resulting from a given set of microstates. Whether these statistics are observed for any real situation depends on how the empirical measure will sample the phase space. \emph{Slow manifolds} are therefore good candidates for relevant restricted phase spaces. Note also that in some cases, the forcing may not be uniform in phase space; in fact, in geophysical flows, it may be relevant to force primarily the slow modes, which provides further justification to a restricted equilibrium approach.
Therefore, in addition to the full phase space canonical probability distribution \eqref{canopdfdefeq}, we shall also compute here the restricted partition function corresponding to the submanifold $\Lambda_0$ (i.e. the slow manifold):
\begin{align}
\rho_0 &= \frac 1 {\partfun_0} e^{-\beta E -\alpha \Gamma_2} \chi_{\Lambda_0},\\
\partfun_0 &= \int_{\Lambda_0} e^{-\beta E -\alpha \Gamma_2} d\mu_{\Lambda_0},
\end{align}
and describe the resulting statistics: average values with respect to the canonical probability distribution $\rho$ will be denoted by brackets $\langle \cdot \rangle$, while average values with respect to the restricted canonical probability distribution $\rho_0$ will be denoted $\langle \cdot \rangle_0$.

\subsection{For Rotating-Stratified flows ($f \neq 0$ and $N \neq 0$)}

\begin{figure}
\centering
\includegraphics{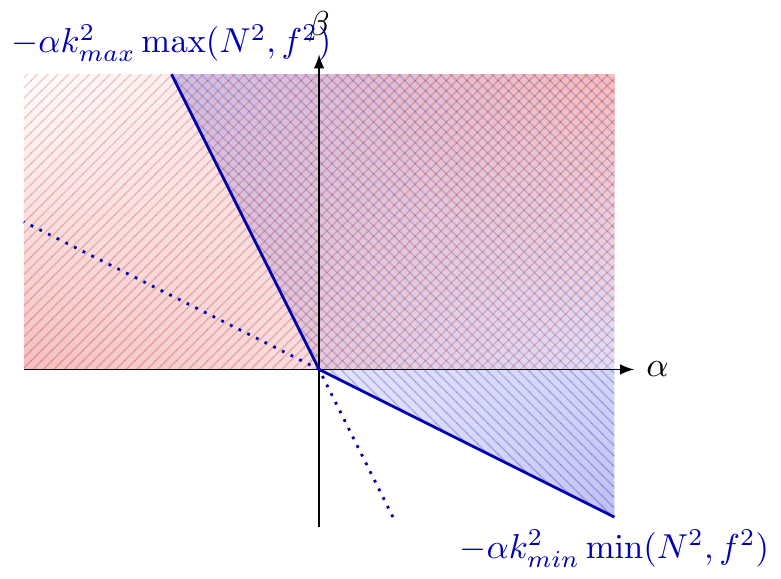}
\caption{(Color online).
Accessible thermodynamic space for rotating-stratified Euler-Boussinesq flows: for the restricted phase space $\Lambda_0$ ($\beta + \alpha k_{min}^2 \min(N^2,f^2)>0$ and $\beta + \alpha k_{max}^2 \max(N^2,f^2) >0$) , 
marked with slanted lines and shaded in blue, and for the restricted phase space $\Lambda_W$ ($\beta>0$), marked with slanted lines and shaded in red. Note that the intersection coincides with the full phase space $\Lambda$. Contrary to 2D turbulence and similarly to 3D turbulence, there is no accessible negative temperature for rotating-stratified Boussinesq flows, but they are recovered in the restricted phase space $\Lambda_0$.}\label{rotstratthermospacefig}
\end{figure}

As mentioned above, the partition function is given by
\begin{align}
\partfun &= \int_\Lambda e^{-\beta E - \alpha \Gamma_2} d\mu_\Lambda,
\intertext{where $d\mu_\Lambda=\prod_{\vec{k} \in \wavevecset{B}} \prod_{i=1}^3 \delta(\hat{k}^i\fourier{u}_i(\vec{k})) d\fourier{u}_i(\vec{k}) d\fourier{\theta}(\vec{k})$ is the measure on the full phase space ($\hat{\vec{k}}=\vec{k}/k$ is the normalized wave vector), so that}
\partfun &= \int_\Lambda e^{-\frac 1 2 \sum_{\vec{k} \in \wavevecset{B}} {^t \vec{X}(\vec{k})^*} \vec{M}_{\beta,\alpha}(\vec{k}) \vec{X}(\vec{k})} d\mu_\Lambda,\\
\partfun &= \prod_{\vec{k} \in \wavevecset{B}} \int e^{- \frac 1 2 {^t\vec{X}(\vec{k})^*} \vec{M}_{\beta,\alpha}(\vec{k}) \vec{X}(\vec{k})} \delta(\hat{k}^i\fourier{u}_i(\vec{k})) d\vec{X}(\vec{k}). \label{partfuneq}
\end{align}
This integral can be evaluated directly by using an integral representation of the Dirac distribution, but it is more enlightening to make use of the description of phase space introduced in \cref{phasespacesec}.
Indeed, we see that expressing the quadratic form in the basis made of the normal modes of the linearized evolution equations $\vec{Z_0}(\vec{k}), \vec{Z_\pm}(\vec{k})$, the partition function --- up to a Jacobian determinant which yields an unimportant constant factor --- reads simply
\begin{align}
\partfun &= \prod_{\vec{k} \in \wavevecset{B}} \int e^{- \frac 1 2 {^t\vec{A}(\vec{k})^*} \vec{\Delta}_{\beta,\alpha}(\vec{k}) \vec{A}(\vec{k})} d\vec{A}(\vec{k}),\\
&= \prod_{\vec{k} \in \wavevecset{B}} \int e^{-\frac 1 2 \Big[ \frac{(\beta+\alpha k^2 \sigma(\vec{k})^2)}{k_\perp^2} \abs{a_0}^2 + \frac{\beta}{k_\perp^2} \abs{a_+}^2 + \frac{\beta}{k_\perp^2} \abs{a_-}^2 \Big]} da_0da_+da_-.
\end{align}
First of all, we see that the \emph{realizability} condition for the Gaussian integral to converge is that $\forall \vec{k} \in \wavevecset{B}, \beta+\alpha k^2 \sigma(\vec{k})^2>0$ and $\beta>0$.
Besides, in this form, it is clear that the partition function is the product of two \emph{restricted partition functions}: on the one hand $\partfun_0$, the partition function restricted to the submanifold $\Lambda_0$ of vortical modes, and on the other hand $\partfun_W$, the partition function restricted to the submanifold $\Lambda_W$ of wave modes. These restricted partition functions are therefore given by:
\begin{align}
\partfun_0 &= \int_{\Lambda_0} e^{-\beta E - \alpha \Gamma_2} d\mu_{\Lambda_0},\\
&= \prod_{\vec{k} \in \wavevecset{B}} \int e^{-\frac 1 2 \frac{(\beta+\alpha k^2 \sigma(\vec{k})^2)}{k_\perp^2} \abs{a_0}^2} da_0,\\
&= \prod_{\vec{k} \in \wavevecset{B}} \sqrt{\frac{2\pi k_\perp^2}{\beta+\alpha (f^2 k_\parallel^2+N^2 k_\perp^2)}},\\
\partfun_W &= \int_{\Lambda_W} e^{-\beta E - \alpha \Gamma_2} d\mu_{\Lambda_W},\label{gwpartfuneq1}\\
&= \prod_{\vec{k} \in \wavevecset{B}} \int e^{-\frac 1 2 \Big[\frac{\beta}{k_\perp^2} \abs{a_+}^2 + \frac{\beta}{k_\perp^2} \abs{a_-}^2 \Big]} da_+da_-,\\
&= \prod_{\vec{k} \in \wavevecset{B}} \sqrt{\frac{(2\pi)^2 k_\perp^2}{\beta^2}},\label{gwpartfuneq3}\\
\partfun &= \partfun_0 \partfun_W,\\
&=\prod_{\vec{k} \in \wavevecset{B}} \sqrt{\frac{(2\pi)^3 k_\perp^4}{\beta^2(\beta+\alpha(f^2 k_\parallel^2+N^2k_\perp^2))}}.
\end{align}
For the Gaussian integrals to converge, it is necessary and sufficient that the quadratic form $\beta E + \alpha \Gamma_2$ (or its restriction to $\Lambda_0$) be positive definite. This means that $\beta>0$ is the condition for $\partfun_W$ to be defined, $\forall \vec{k} \in \wavevecset{B}, \beta+\alpha(f^2 k_\parallel^2 + N^2 k_\perp^2)>0$ --- this is equivalent to $\beta+\alpha k_{min}^2 \min(N^2,f^2)$ for $\alpha>0$ and to $\beta+\alpha k_{max}^2 \max(N^2,f^2)>0$ when $\alpha<0$ --- is the condition for $\partfun_0$ to be defined, and as mentioned above, $\partfun$ requires both conditions. We illustrate these conditions in \Cref{rotstratthermospacefig}.
It is worthy of note that it is the presence of the inertia-gravity waves modes which imposes the $\beta>0$ condition for the full phase space. In other words, Euler-Boussinesq flows cannot reach absolute equilibrium states of negative temperatures, similarly to 3D homogeneous isotropic flows. Nevertheless, the system with phase space restricted to the vortical modes can reach such negative temperature states. In that case, we can identify three regimes by analogy with 2D flows~\citep{Kraichnan1975,Kraichnan1980} and 3D helically constrained flows~\citep{Herbert2013d}:
\begin{enumerate}[label={(\Roman*) }  ]
\item $\alpha >0$, $\beta < 0$
\item $\alpha > 0$, $\beta > 0$
\item $\alpha <0$, $\beta>0$
\end{enumerate}

Now, let us compute the average energy and potential enstrophy spectra at statistical equilibrium.
The average energy at equilibrium for the full and restricted phase space read, respectively:
\begin{align}
\langle E \rangle &= \int_\Lambda E \rho d\mu_{\Lambda}, & \langle E \rangle_0 &= \int_{\Lambda} E \rho_0 d\mu_\Lambda,\\
& = - \frac{\partial \ln \partfun}{\partial \beta}, & &= - \frac{\partial \ln \partfun_0}{\partial \beta},\\
& = \frac 1 2 \sum_{\vec{k} \in \wavevecset{B}} \Big[ \frac 2 \beta + \frac{1}{\beta+\alpha(f^2 k_\parallel^2+N^2k_\perp^2)}\Big], && = \frac 1 2 \sum_{\vec{k} \in \wavevecset{B}} \frac{1}{\beta+\alpha(f^2 k_\parallel^2+N^2k_\perp^2)}.
\intertext{We see that the wave modes reach energy equipartition at statistical equilibrium, as expected since they do not contribute to the potential vorticity $\Gamma_2$. Replacing the discrete sums by integrals, we have}
\langle E \rangle & = \frac 1 2 \int \Big[ \frac 2 \beta + \frac{1}{\beta+\alpha(f^2 k_\parallel^2+N^2k_\perp^2)}\Big] d\vec{k}, & \langle E \rangle_0& = \frac 1 2 \int \frac{1}{\beta+\alpha(f^2 k_\parallel^2+N^2k_\perp^2)}d\vec{k}.
\end{align}
We introduce the anisotropic energy spectrum $e(k_\perp,k_\parallel)$ such that $E = \underset{k_{min}^2 \leq k_\perp^2+k_\parallel^2 \leq k_{max}^2}{\int \int} e(k_\perp,k_\parallel) dk_\perp dk_\parallel$. We have
\begin{align}
\langle e(k_\perp,k_\parallel) \rangle &= \pi k_\perp \Big[ \frac 2 \beta + \frac{1}{\beta+\alpha(f^2 k_\parallel^2+N^2k_\perp^2)}\Big], &  \langle e(k_\perp,k_\parallel) \rangle_0 & = \frac{\pi k_\perp}{\beta+\alpha(f^2 k_\parallel^2+N^2k_\perp^2)}.
\end{align}
We also introduce the isotropic energy spectrum $E(k)$, to allow for comparison with the isotropic cases: $E=\int_{k_{min}}^{k_{max}} E(k)dk$, so that
\begin{align}
\langle E(k) \rangle &= \int_0^{\pi/2} \langle e(k \cos \phi, k \sin \phi) \rangle kd\phi, & \langle E(k) \rangle_0 &= \int_0^{\pi/2} \langle e(k \cos \phi, k \sin \phi) \rangle_0 kd\phi
\end{align}
The integration yields
\begin{align}
\langle E(k)\rangle_0 &=
\begin{cases}
\frac{\pi k}{\sqrt{\alpha(f^2-N^2)(\beta+\alpha N^2 k^2)}} \arctan \Big( k \sqrt{\frac{\alpha (f^2-N^2)}{\beta+\alpha N^2 k^2}}\Big) & \text{if } f > N,\\
\frac{\pi k^2}{\beta+\alpha N^2 k^2} & \text{if } f=N,\\
\frac{\pi k}{2\sqrt{\alpha(N^2-f^2)(\beta+\alpha N^2 k^2)}} \ln \Big( \frac{\sqrt{\beta+\alpha N^2 k^2}+k\sqrt{\alpha(N^2-f^2)}}{\sqrt{\beta+\alpha N^2 k^2}-k\sqrt{\alpha(N^2-f^2)}} \Big) & \text{if } N > f,
\end{cases}
\end{align}
and we have $\langle E(k) \rangle = \langle E(k) \rangle_0 + 2\pi k^2/\beta$. Note that the equilibrium energy spectrum depends continuously on $f$ and $N$.
We can similarly obtain the average potential enstrophy:
\begin{align}
\langle \Gamma_2 \rangle & = \langle \Gamma_2 \rangle_0 = - \frac{\partial \ln \partfun_0}{\partial \alpha},\\
& = \frac 1 2 \sum_{\vec{k} \in \wavevecset{B}} \frac{f^2k_\parallel^2+N^2k_\perp^2}{\beta+\alpha(f^2 k_\parallel^2+N^2k_\perp^2)},
\end{align}
so that introducing the anisotropic potential enstrophy spectrum $\gamma_2(k_\perp,k_\parallel)$ and the isotropic potential enstrophy spectrum $\Gamma_2(k)$ as above, we have,
\begin{align}
\langle \gamma_2(k_\perp,k_\parallel) \rangle_0 &= \frac{\pi k_\perp (f^2 k_\parallel^2+N^2 k_\perp^2)}{\beta+\alpha (f^2 k_\parallel^2+N^2 k_\perp^2)},\\
\langle \Gamma_2(k) \rangle_0 &= \frac{\pi k^2}{\alpha} - \frac{\beta}{\alpha} \langle E(k) \rangle_0,
\end{align}
where the last equality can also be interpreted as equipartition of the general invariant $\beta E + \alpha \Gamma_2$ on the manifold $\Lambda_0$; this is analogous to 2D turbulence, except that we are now in 3D, hence the $k^2$ instead of the $k$.
\begin{figure}
\includegraphics[width=\textwidth]{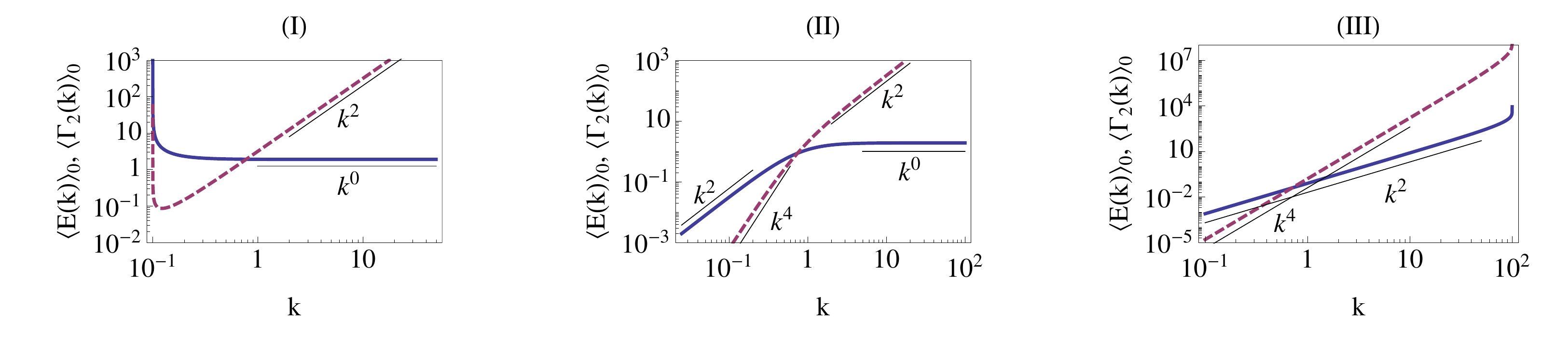}
\caption{(Color online). Isotropic spectra of energy $E$ (solid blue lines) and potential enstrophy $\Gamma_2$ (dashed red) for rotating and stratified flows (here in the case $f>N$) at statistical equilibrium restricted to the submanifold of vortical modes $\Lambda_0$, in the different $(\alpha,\beta)$ regimes. Left: $\alpha>0, \beta <0$ (regime I); the spectra have an infrared divergence corresponding to spectral condensation. 
Middle: $\alpha>0, \beta>0$ (regime II); intermediate regime, the spectra increase with $k$ but the energy spectrum is flat at large $k$. Right: $\alpha<0,\beta>0$ (regime III); the spectra increase as $k$ increases, and there is an ultraviolet divergence.}\label{rotstratvortspecfig}
\end{figure}
The isotropic spectra for energy and potential enstrophy in the restricted canonical ensemble are shown in \cref{rotstratvortspecfig} for the three different regimes described above. The qualitative behavior is the same as for 2D turbulence~\citep{Kraichnan1980}, but the scalings are different: at small $k$ the energy scales like $k^2$ (energy equipartition in 3D), while at large $k$, it has a flat spectrum corresponding to potential enstrophy equipartition. In the energy dominated regime --- regime (I) --- there is an infrared divergence, which indicates that a condensation of energy in the gravest modes should be expected. This hints at the presence of an inverse cascade for the vortical modes, in a similar way to the inverse cascade of 2D turbulence.
The potential enstrophy dominated regime --- regime (III) --- is similar to 3D HIT, with potential enstrophy playing the role of helicity.
Only the case $f>N$ is shown on the figure, but the cases $f=N$ and $f<N$ have a similar behavior.
\begin{figure}
\includegraphics[width=\textwidth]{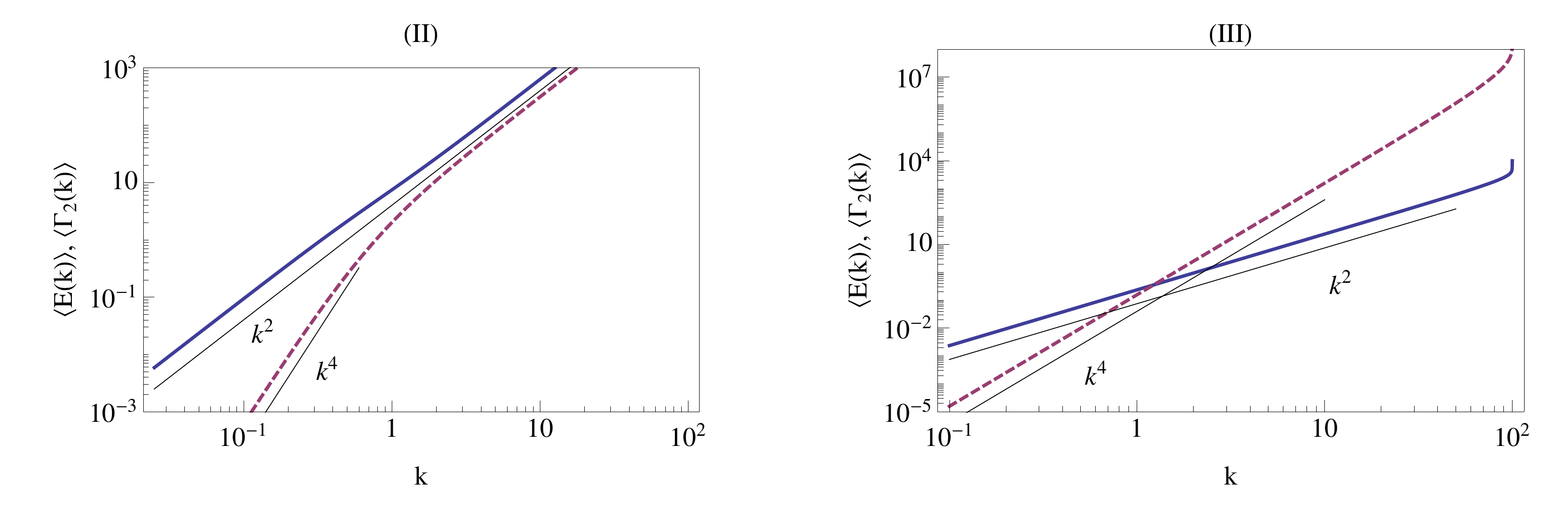}
\caption{(Color online). Isotropic spectra of energy $E$ (solid blue) and potential enstrophy $\Gamma_2$ (dashed red) for rotating and stratified flows (here in the case $f>N$) at absolute equilibrium, in the different $(\alpha,\beta)$ regimes. 
Left: $\alpha>0, \beta>0$ (regime II); the energy spectrum is almost in equipartition, the potential enstrophy spectrum increases with scalings $k^4$ at low-$k$ and $k^2$ (equipartition) at large $k$. Right: $\alpha<0,\beta>0$ (regime III); the spectra increase as $k$ increases, with energy equipartition, and there is an ultraviolet divergence.}\label{rotstratspecfig}
\end{figure}
By contrast, on the full phase space, the presence of inertia-gravity waves prevents access to regime (I); no condensation of energy at small scales is expected. The potential enstrophy dominated regime, with energy equipartition and ultraviolet divergence --- regime (III) --- remains the same as above. In the intermediate regime, the energy spectrum at small scales is dominated by the waves and instead of the flat spectrum coming from potential enstrophy equipartition, we now see the wave modes equipartition regime with a $k^2$ scaling. 
In all cases, and similarly to 3D HIT, the energy spectrum is close to equipartition. However, at variance with 3D HIT, in the small scales of the intermediate regime --- regime (II) --- potential enstrophy also reaches equipartition. This is not in contradiction with energy equipartition because the former is due to the vortical modes reaching equipartition, while the latter is due to equipartition of the waves mode.

\subsection{For Purely Stratified flows ($f=0$)}

In the absence of rotation, the invariants of the system remain the same, and the partition function can be obtained simply by setting $f=0$ in the partition function of rotating-stratified flows. Clearly, the partition function $\partfun_W$ for the gravity wave modes remain the same (see \eqref{gwpartfuneq1}-\eqref{gwpartfuneq3}), while the partition function $\partfun_0$ for the vortical modes becomes
\begin{align}
\partfun_0 &= \prod_{\vec{k} \in \wavevecset{B}} \sqrt{\frac{2\pi k_\perp^2}{\beta+\alpha N^2 k_\perp^2}},\\
\intertext{so that the full partition function becomes}
\partfun &= \prod_{\vec{k} \in \wavevecset{B}} \sqrt{\frac{(2\pi)^3 k_\perp^4}{\beta^2(\beta+\alpha N^2k_\perp^2)}}.
\end{align}
\begin{figure}
\centering
\includegraphics{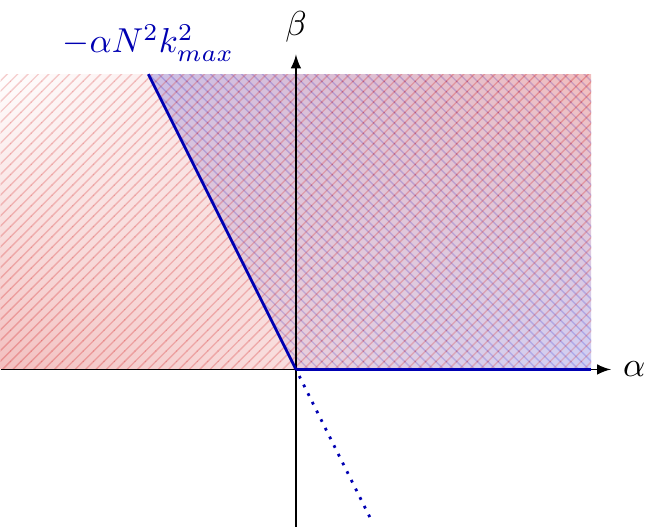}
\caption{(Color online).
Accessible thermodynamic space for purely stratified Euler-Boussinesq flows: for the restricted phase space $\Lambda_0$ ($\beta >0$ and $\beta + \alpha N^2 k_{max}^2 >0$) , 
marked with slanted lines and shaded in blue, and for the restricted phase space $\Lambda_W$ ($\beta>0$), marked with slanted lines and shaded in red. Note that the intersection coincides with the full phase space $\Lambda$. Contrary to 2D turbulence and similarly to 3D turbulence, there is no accessible negative temperature for purely stratified Boussinesq flows, and unlike the rotating case, they are not recovered in the restricted phase space $\Lambda_0$.}\label{stratthermospacefig}
\end{figure}
The realizability condition for the restricted partition function is that $\forall \vec{k}, \beta + \alpha N^2 k_\perp^2 >0$. In principle, there exist wave vectors $\vec{k}$ with vanishing horizontal component, therefore the condition implies $\beta >0$, as well as $\beta+\alpha N^2 k_{max}^2>0$.
In the full phase space, the condition $\beta > 0$ is enforced from the start. The space of accessible thermodynamic parameters $(\alpha,\beta)$ is shown in \cref{stratthermospacefig}.

The average energy at statistical equilibrium is given by:
\begin{align}
\langle E \rangle & = \frac 1 2 \sum_{\vec{k} \in \wavevecset{B}} \Big[ \frac 2 \beta + \frac{1}{\beta+\alpha N^2k_\perp^2}\Big], & \langle E \rangle_0& = \frac 1 2 \sum_{\vec{k} \in \wavevecset{B}} \frac{1}{\beta+\alpha N^2k_\perp^2},
\intertext{so that the anisotropic energy spectra are}
\langle e(k_\perp,k_\parallel) \rangle &= \pi k_\perp \Big[ \frac 2 \beta + \frac{1}{\beta+\alpha N^2k_\perp^2}\Big], &  \langle e(k_\perp,k_\parallel) \rangle_0 & = \frac{\pi k_\perp}{\beta+\alpha N^2k_\perp^2}, 
\intertext{and the isotropic energy spectra}
\langle E(k) \rangle &= \langle E(k) \rangle_0 + \frac{2\pi k^2}{\beta}, & \langle E(k) \rangle_0 &= \frac{\pi k}{2\sqrt{\alpha N^2(\beta+\alpha N^2 k^2)}} \ln \Big( \frac{\sqrt{\beta+\alpha N^2 k^2}+kN\sqrt{\alpha}}{\sqrt{\beta+\alpha N^2 k^2}-kN\sqrt{\alpha}} \Big).
\end{align}
Similarly, the anisotropic and isotropic spectra of potential enstrophy at statistical equilibrium are given by
\begin{align}
\langle \gamma_2(k_\perp,k_\parallel) \rangle_0 &= \frac{\pi N^2 k_\perp^3}{\beta+\alpha N^2 k_\perp^2},\\
\langle \Gamma_2(k) \rangle_0 &= \frac{\pi k^2}{\alpha} - \frac{\beta}{\alpha} \langle E(k) \rangle_0,
\end{align}
similarly to the case with rotation.
Hence, stratified turbulence in the restricted canonical ensemble features the two regimes (II) and (III) with the same qualitative behavior as in the presence of rotation (\cref{rotstratvortspecfig}). In the full canonical ensemble as well, the system behaves qualitatively like the two regimes (II) and (III) in the presence of rotation (\cref{rotstratspecfig}).
In particular, in the absence of rotation, there is in all rigor no negative temperature regime, which could indicate an inverse cascade of energy, even in the restricted ensemble. This regime can be recovered by imposing an infrared cutoff in $k_\perp$, but this is somewhat artificial.

\subsection{For Purely Rotating flows ($N=0$)}\label{rotmecastatsec}

The restricted partition function is given by
\begin{align}
\partfun_{2D} &= \int_{\Lambda_{2D}} e^{-\beta E_{2D} -\beta_\parallel E_\parallel - \alpha \Gamma_2} d\mu_{\Lambda_{2D}},\\
&= \prod_{\vec{k} \in \wavevecset{B}_{2D}} \int e^{-\frac 1 2 \left[ \frac{\beta_\parallel}{k_\perp^2} \abs{D}^2 + \left(\frac{\beta}{k_\perp^2}+\alpha\right) \abs{\zeta}^2 \right]} dD d \zeta,\\
&= \prod_{\vec{k} \in \wavevecset{B}_{2D}} \sqrt{\frac{(2\pi)^2k_\perp^4}{\beta_\parallel(\beta+\alpha k_\perp^2)}}.
\end{align}
The realizability condition reads $\beta_\parallel >0$ and $\forall \vec{k} \in \wavevecset{B}_{2D}, \beta+\alpha k^2>0$, which amounts to $\beta+\alpha k_{min}^2>0$ and $\beta+\alpha k_{max}^2>0$. In particular, states of negative temperature $\beta$ can be attained. The accessible thermodynamic space is the same as in \cref{rotstratthermospacefig} with $N=f$, up to a rescaling of $\alpha$.
In particular, the three regimes identified for the slow modes of the rotating-stratified case exist here as well.
The average energy at statistical equilibrium in the restricted ensemble is
\begin{align}
\langle E_{2D} \rangle_{2D} &= - \frac{\partial \ln \partfun_{2D}}{\partial \beta}, & \langle E_\parallel \rangle_{2D} &= - \frac{\partial \ln \partfun_{2D}}{\partial \beta_\parallel}, \\
&= \frac 1 2 \sum_{\vec{k} \in \wavevecset{B}_{2D}} \frac{1}{\beta+\alpha k_\perp^2}, & &=\frac 1 2 \sum_{\vec{k} \in \wavevecset{B}_{2D}} \frac{1}{\beta_\parallel}.
\end{align}
The anisotropic and isotropic energy spectrum at statistical equilibrium in the restricted ensemble are therefore given by
\begin{align}
\langle e_{2D}(k_\perp,k_\parallel) \rangle_{2D} &= \frac{\pi k_\perp}{\beta+\alpha k_\perp^2} \delta(k_\parallel), & \langle e_\parallel(k_\perp,k_\parallel) \rangle_{2D} &= \frac{\pi k_\perp}{\beta_\parallel} \delta(k_\parallel),\\
\langle E_{2D}(k) \rangle_{2D} &= \frac{\pi k^2}{\beta+\alpha k^2}, & \langle E_{\parallel}(k) \rangle_{2D} &= \frac{\pi k^2}{\beta_\parallel}.
\end{align}
Similarly, we obtain the anisotropic and isotropic spectra for enstrophy:
\begin{align}
\langle \gamma_2^{2D}(k_\perp,k_\parallel) \rangle_{2D} &= \frac{\pi k_\perp^3}{\beta+\alpha k_\perp^2} \delta(k_\parallel),\\
\langle \Gamma_2^{2D}(k) \rangle_{2D} &= \frac{\pi k^4}{\beta+\alpha k^2}.
\end{align}
We see that the average isotropic energy spectrum for the slow modes of rotating turbulence in the restricted ensemble ($\langle E_{2D}(k) \rangle_{2D}$) is similar to that of the slow modes in rotating-stratified turbulence with $N=f$, up to a rescaling of the $\alpha$ parameter. In particular, the qualitative behavior and the scalings for the three regimes are the same as in \cref{rotstratvortspecfig}, with an infrared divergence of the spectra in the negative temperature regime.

\section{Discussion of the Hypotheses}\label{discussec}

\subsection{How slow is the slow manifold?}\label{slowdiscussec}

The slow modes/fast waves decomposition was introduced here through the normal modes of the linearized dynamics in Fourier space. 
Still in the linearized framework, we have seen that in the presence of rotation and stratification, the slow modes satisfy the balance relations and are the modes which carry potential vorticity (see \cref{phasespacesec} for the subtlety of the purely stratified case). In fact, for the full, nonlinear dynamics, there is no guarantee that these two properties still coincide, although one may still introduce a wave-vortical modes decomposition \citep{Staquet1989}, or define slow manifolds and \emph{superbalance} relations \citep{Vanneste2013}.
Here, there is no consequence on the computations we make, because the decomposition is only introduced as a way to describe phase space. Neither the linear nor the exact nonlinear dynamics were used to compute the partition functions, as the absolute or restricted equilibrium probability densities only depend on the invariants of the system. 
However, there is a consequence on the interpretation of the restricted ensemble. Indeed, for the statistics predicted by the restricted probability density to describe accurately the real system, it is necessary that it should not wander too far away from the submanifold $\Lambda_0$. One justification for this comes from an argument of time scale separation. By definition, the vortical mode have zero linear frequency; their dynamics is trivial at the linear level. In fact, their dynamics is governed by nonlinear advection, with a characteristic timescale given by the eddy turnover time $\tau_0 = L/U$, where $L$ and $U$ are a characteristic length scale and a characteristic velocity. On the other hand, the timescale of inertia gravity waves is given by the inverse of the frequency $\sigma(\vec{k})$, which is bounded: $\min(f,N) \leq \sigma(\vec{k}) \leq \max(f,N)$. Therefore, the ratio of the two timescales satisfies
\begin{align}
\epsilon = \frac{\sigma(\vec{k})^{-1}}{\tau_0} &\leq \frac{U}{L\min(f,N)},\\
&\leq \max(\Ro, \Fr),
\end{align}
where $\Ro$ and $\Fr$ are respectively the Rossby and Froude numbers. 
It follows from an argument of optimal truncation in the asymptotic series defining the slow manifolds that the accuracy of the balanced model is exponential in $\epsilon$. In other words, due to the fast inertia-gravity waves, the system oscillates around the slow manifold, thereby defining a \emph{fuzzy manifold} of exponentially small width as $\epsilon$ decreases \citep{Lorenz1987a,Warn1997}.
Typically, for the large scales of the atmosphere and the ocean, we have $\Fr \ll \Ro \ll 1$. Therefore the time scales are well separated: $\epsilon \ll 1$, the width of the \emph{fuzzy manifold} is small, and we may expect the restricted ensemble predictions to hold. In particular, there should be an inverse cascade, due to the vortical modes, in such cases.

At smaller scale (e.g. mesoscales in the atmosphere and sub-mesoscales in the ocean), we may still have $\Fr \ll 1$, but $\Ro$ becomes of order one. When this happens, the argument above breaks down; there is no longer a clear time scale separation. As we have noted above, the slow modes of purely stratified turbulence do not lead to an inverse cascade but rather to a direct cascade, at least according to the statistical mechanics argument --- and numerical simulations support this view. The transition to stratified turbulence with small rotation is probably smooth due to the fact that the restricted ensemble prediction for rotating/stratified turbulence breaks down when the Rossby number is of order one or larger.

The case of purely rotating turbulence is different. In this case again, there is no clear separation of time scales. The frequencies of the inertial waves belong to the interval $[0,f]$, in the limit of a large domain $L \to \infty$. The ratio $\epsilon$ is only bounded from below: $\epsilon \ge \Ro$. However, we know that for strong enough rotation, the flow tends to form columns, which corresponds to two-dimensional ($k_\parallel=0$) modes \citep{Taylor1917,Proudman1916,LSmith1999b}. This goes in support of the relevance of the restricted ensemble predictions for rotating flows.

\subsection{How quadratic is potential enstrophy?}\label{pvdiscussec}

Another hypothesis that we have made here is that the potential enstrophy is well approximated by retaining only the quadratic part, instead of the full expression which involves cubic and quartic terms. 
Here, the motivation is mainly technical: a quadratic potential enstrophy ensures that the canonical distribution (absolute or restricted) is a Gaussian, and therefore that the partition function can be computed analytically.
In fact, the issue is deeper than that and non-quadratic potential enstrophy may have peculiar consequences; \cite{Herring1994} have for instance reported generation of potential enstrophy at large scales by viscosity, at variance with the usual small scale dissipation role of viscosity. 

We can expect this approximation to perform well in the limit of strong rotation and stratification, as quasi-geostrophic turbulence has a quadratic potential enstrophy \citep{Charney1971}. 

With no or weak rotation, scale analysis shows that the non-quadratic terms are small when the vertical Froude number $\Fr_v = U/HN$, where $H$ is the vertical length scale, is small: $Fr_v \ll 1$ \citep{Waite2006}. But as argued by \cite{Billant2001} on the basis of an invariance property of the inviscid equations, the vertical Froude number always remains of order one: $\Fr_v = O(1)$, even when the horizontal Froude number $Fr_h=U/LN$ is small. This means that the width of the layers which form in strongly stratified flows is given by the buoyancy scale $L_b=U/N$, as observed in numerical simulations~\citep{Riley2003,Lindborg2006,Waite2011}.
However, in the numerical simulations of \cite{Aluie2011}, for strongly stratified and weakly rotating flows, potential enstrophy is well approximated by the quadratic terms. As suggested by \cite{Waite2013}, this may be due to the effect of viscosity: when viscous effects become strong enough, the argument of \cite{Billant2001} breaks down. A viscous coupling between the layers sets in, and the vertical length scale is no longer given by the buoyancy scale but rather by the viscous scale $\sqrt{\nu L_h/U}$ \citep{Riley2003,GodoyDiana2004,Brethouwer2007} and the vertical Froude number can reach smaller values. \cite{Brethouwer2007} and \cite{Waite2013} suggested to discriminate the cases by the buoyancy Reynolds number $\Re_b = \Fr_h^2 \Re$. When $\Re_b$ is small (viscously coupled regime), potential enstrophy is well approximated by its quadratic part, while for larger $\Re_b$, the higher order terms are not negligible.

\section{Conclusion}

In this paper, we have provided new theoretical arguments to study the possibility of inverse or direct cascades of energy in rotating and/or stratified flows.
Using a description of phase space in terms of slow eddy and fast wave modes, based on the normal modes of the linearized equations of motion~\citep{Leith1980,Bartello1995}, we have investigated the role of both kind of modes through statistical mechanics arguments. We have adapted the idea of restricted partition functions introduced in condensed matter to study metastability; here, the integral defining the partition function is restricted to the slow manifold. 
We find that for rotating and stratified or purely rotating flows, negative temperature states are possible, for which the energy spectrum at restricted equilibrium has an infrared divergence characteristic of an inverse cascade regime. At absolute equilibrium (i.e. not restricting the integral defining the partition function), the presence of waves results in the disappearance of the negative temperature regime, and therefore a direct cascade of energy should be expected in agreement with \cite{Bartello1995}. For purely stratified flows, negative temperature states do not exist, even in the restricted ensemble (see also \cite{Waite2004} for the absolute equilibrium case), because of the shear modes ($k_\perp=0$). Excluding these modes, we would recover a possibility for an inverse cascade, but this does not seem justified as the shear modes are part of the slow manifold and play an important role in the dynamics, like in the thermal wind.

The results above rely on the assumptions that potential enstrophy is quadratic and that the system remains in the vicinity of the slow manifold. Both these conditions should be satisfied when rotation and stratification are strong, or for purely rotating flows. They begin to break down when rotation becomes weaker.

These results are consistent with the findings of numerical simulations of rotating and stratified flows, which tend to show that stratified flows exhibit an inverse cascade of energy only when rotation is strong enough~\citep[e.g.][]{Marino2013b}. They indicate that the largest scales of the atmosphere and ocean should have an inverse cascade of energy, due to the vortical modes, in accordance with the theory of quasi-geostrophic turbulence~\citep{Charney1971,Rhines1979}, while at smaller scales, when the effect of rotation weakens (e.g. in the atmospheric mesoscale or oceanic sub-mesoscale), the energy transfers should be downscale. Such a dual cascade was indeed observed recently in direct numerical simulations~\citep{Pouquet2013}.

\appendix
\section{Notes on the normal modes of the linearized Boussinesq equations}\label{normalmodessec}

In addition to the two sets of variables $\vec{X}$ and $\vec{Z}$ considered in \cref{phasespacesec}, we introduce
\begin{align}
\vec{Y}(\vec{k})&=\begin{pmatrix} \zeta(\vec{k}) \\ \delta(\vec{k}) \\ \fourier{\theta}(\vec{k})\end{pmatrix},& \vec{L'}(\vec{k})=\begin{pmatrix}
0					& -f 					& 0	 							\\
f \frac{k_\parallel^2}{k^2}	& 0					& i N k_\parallel \frac{k_\perp^2}{k^2} 	\\
0					&i \frac{N}{k_\parallel}	& 0	 							\\
\end{pmatrix},
\intertext{so that}
\dot{\vec{Y}}(\vec{k}) &= \vec{L'}(\vec{k})\vec{Y}(\vec{k}).&
\end{align}
The various descriptions are related by linear transformations:
\begin{align}
\vec{Y}(\vec{k}) &= \vec{Q}(\vec{k}) \vec{S} \vec{X}(\vec{k}), & \vec{Z}(\vec{k}) &= \vec{R}(\vec{k})\vec{Y}(\vec{k}),
\intertext{with}
\vec{Q}(\vec{k}) &= \begin{pmatrix}
i k_1		& ik_2	& 0\\
-i k_2	& ik_1	& 0\\
0		& 0		& 1
\end{pmatrix},
&
\vec{R}(\vec{k})&=\begin{pmatrix}
1	& 0					& 0\\
0	& \frac{k}{k_\parallel}		& 0\\
0	& 0					& -k_\perp
\end{pmatrix},\\
\vec{S} &= \begin{pmatrix}
1 & 0 & 0 & 0\\
0 & 1 & 0 & 0\\
0 & 0 & 0 &1
\end{pmatrix}.&&&&
\end{align}
These linear transformations have physical interpretations: the matrix $\vec{Q}(\vec{k})$ is the product of a unitary transformation $\vec{U} \in U(3)$ and two dilatations $\vec{D}_1(k_\perp)$ and $\vec{D}_2(k_\perp)$;
the matrix $\vec{R}(\vec{k})$ can be seen as the product of two dilatations $\vec{D}_2(k/k_\parallel)$ and $\vec{D}_3(-k_\perp)$: 
\begin{align}
\vec{Q}(\vec{k})&=\vec{U} \vec{D}_1(k_\perp)\vec{D}_2(k_\perp),&&\\
\vec{R}(\vec{k})&=\vec{D}_2(k/k_\parallel)\vec{D}_3(-k_\perp),&&
\intertext{with}
\vec{U}&=\begin{pmatrix} i \frac{k_1}{k_\perp} & i\frac{k_2}{k_\perp} & 0\\ -i\frac{k_2}{k_\perp} & i \frac{k_1}{k_\perp} & 0 \\ 0 & 0 & 1\end{pmatrix},& \vec{D}_i(x)&=\vec{I}_3+(x-1)\vec{E}_{ii}.
\end{align}

We have $\det \vec{Q}(\vec{k})=-k_\perp^2$ and $\det \vec{R}(\vec{k})=k k_\perp/k_\parallel$.

Note that the matrices $\vec{R}(\vec{k})$ and $\vec{L'}(\vec{k})$ are not defined when $k_\parallel=0$. As explained in the text, for the modes with $k_\parallel=0$, $\delta(\vec{k})$ always vanishes and is therefore not an appropriate dynamical variable. Hence in that case one should use the vector $\vec{Z}(\vec{k})$ rather than $\vec{Y}(\vec{k})$.

\end{document}